\documentclass[sigconf]{aamas}

\usepackage{balance} 
\usepackage{amsmath}
\usepackage{tikz}
\usepackage{mathtools}
\usepackage{amsthm}
\usepackage{microtype}
\usepackage{graphicx}
\usepackage{subfigure}
\usepackage{booktabs} 
\usepackage{makecell}
\usepackage{hyperref} 

\makeatletter
\gdef\@copyrightpermission{
  \begin{minipage}{0.2\columnwidth}
   \href{https://creativecommons.org/licenses/by/4.0/}{\includegraphics[width=0.90\textwidth]{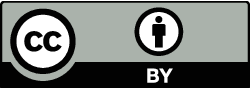}}
  \end{minipage}\hfill
  \begin{minipage}{0.8\columnwidth}
   \href{https://creativecommons.org/licenses/by/4.0/}{This work is licensed under a Creative Commons Attribution International 4.0 License.}
  \end{minipage}
  \vspace{5pt}
}
\makeatother

\setcopyright{ifaamas}
\acmConference[AAMAS '25]{Proc.\@ of the 24th International Conference
on Autonomous Agents and Multiagent Systems (AAMAS 2025)}{May 19 -- 23, 2025}
{Detroit, Michigan, USA}{Y.~Vorobeychik, S.~Das, A.~Nowé  (eds.)}
\copyrightyear{2025}
\acmYear{2025}
\acmDOI{}
\acmPrice{}
\acmISBN{}

\acmSubmissionID{357}

\title[Learning Task Decompositions]{Learning Symbolic Task Decompositions for Multi-Agent Teams}

\author{Ameesh Shah}\authornote{Equal Contribution}
\affiliation{
  \institution{UC Berkeley}
  \city{Berkeley}
  \country{USA}}
\email{ameesh@berkeley.edu}

\author{Niklas Lauffer}\authornotemark[1]
\affiliation{
  \institution{UC Berkeley}
  \city{Berkeley}
  \country{USA}}
\email{nlauffer@berkeley.edu}

\author{Thomas Chen}\authornotemark[1]
\affiliation{
  \institution{UC Berkeley}
  \city{Berkeley}
  \country{USA}}
\email{thomasychen@berkeley.edu}

\author{Nikhil Pitta}\authornotemark[1]
\affiliation{
  \institution{UC Berkeley}
  \city{Berkeley}
  \country{USA}}
\email{nikhil.pitta@berkeley.edu}

\author{Sanjit A. Seshia}
\affiliation{
  \institution{UC Berkeley}
  \city{Berkeley}
  \country{USA}}
\email{sseshia@berkeley.edu}

\begin{abstract}
    One approach for improving sample efficiency in cooperative multi-agent learning is to decompose overall tasks into sub-tasks that can be assigned to individual agents.   
    We study this problem in the context of \textit{reward machines}: symbolic tasks that can be formally decomposed into sub-tasks. In order to handle settings without \textit{a priori} knowledge 
    of the environment, we introduce a framework that can learn the optimal decomposition from model-free interactions with the environment. Our method uses a task-conditioned architecture to simultaneously learn an optimal decomposition and the corresponding agents' policies for each sub-task.
    In doing so, we remove the need for a human to manually design the optimal decomposition while maintaining the sample-efficiency benefits of improved credit assignment.  
    We provide experimental results in several deep reinforcement learning settings, demonstrating the efficacy of our approach. Our results indicate that our approach succeeds even in environments with codependent agent dynamics, enabling synchronous multi-agent learning not achievable in previous works.\footnote{Code is available at \href{https://github.com/thomasychen/LOTaD}{https://github.com/thomasychen/LOTaD}}
\end{abstract}


\newcommand{\BibTeX}{\rm B\kern-.05em{\sc i\kern-.025em b}\kern-.08em\TeX}

\begin{document}


\pagestyle{fancy}
\fancyhead{}


\maketitle 

\newcommand{\savespace}[1]{\textcolor{red}{!}}

\newcommand{\ameesh}[1]{{\footnotesize\color{red}[{\bf Ameesh:} \textsf{#1}]}}
\newcommand{\nik}[1]{{\footnotesize\color{purple}[{\bf Nik:} \textsf{#1}]}}
\newcommand{\thomas}[1]{{\footnotesize\color{orange}[{\bf Thomas:} \textsf{#1}]}}
\newcommand{\nikhil}[1]{{\footnotesize\color{orange}[{\bf Nikhil:} \textsf{#1}]}}

\newcommand{\AP}{\text{AP}}
\newcommand{\mdp}{\mathcal{M}}
\newcommand{\markovgame}{\mathcal{G}}
\newcommand{\rmstates}{U}
\newcommand{\states}{\mathcal{S}}
\newcommand{\buchiaccepts}{\buchistates^*}
\newcommand{\actions}{\mathcal{A}}
\newcommand{\taskspec}{\varphi}
\newcommand{\rewmach}{\mathcal{R}}
\newcommand{\labelingfunction}{L^\mdp}
\newcommand{\productmdp}{\mdp^{\taskspec}}
\newcommand{\decompositions}{\mathcal{D}}
\newcommand{\sampletau}{\tau \sim \mdp^{\varphi}_{\pi}}

\section{Introduction}

Using a single reward signal for a team of agents in multi-agent reinforcement learning (MARL) can make it challenging for agents to understand how their individual behavior impacts the overall reward. This challenge in MARL is known as the \textit{credit assignment problem} \cite{agogino2004creditassignment} and can severely limit the effectiveness of naive reinforcement learning approaches in the multi-agent setting \cite{Oroojlooy2023cmarlsurvey, sunehag2017value}.

One method for addressing the credit assignment problem is to formulate the task as a \textit{symbolic concept} that can be precisely decomposed into sub-tasks for assignment to individual agents~\cite{neary2020reward, smith2023automatic}. By using the reward signal from each sub-task, agents are credited for completing the specific sub-task they are assigned, even if the overall task is not achieved. 

Previous literature has established sufficient conditions for ``valid" decompositions of multi-agent \textit{reward machines}~\cite{Icarte2020RewardMachine}, an automaton-based symbolic task structure, where satisfaction of the sub-tasks provably satisfies the overall task \cite{neary2020reward}. Human-designed decompositions that satisfy these validity conditions help the agents learn to accomplish the task more quickly in tabular settings. Further work demonstrates that valid decompositions can be automatically generated based on human-designed heuristics \cite{smith2023automatic}.

\begin{figure}[t]
\centering
\includegraphics[width=1\columnwidth]{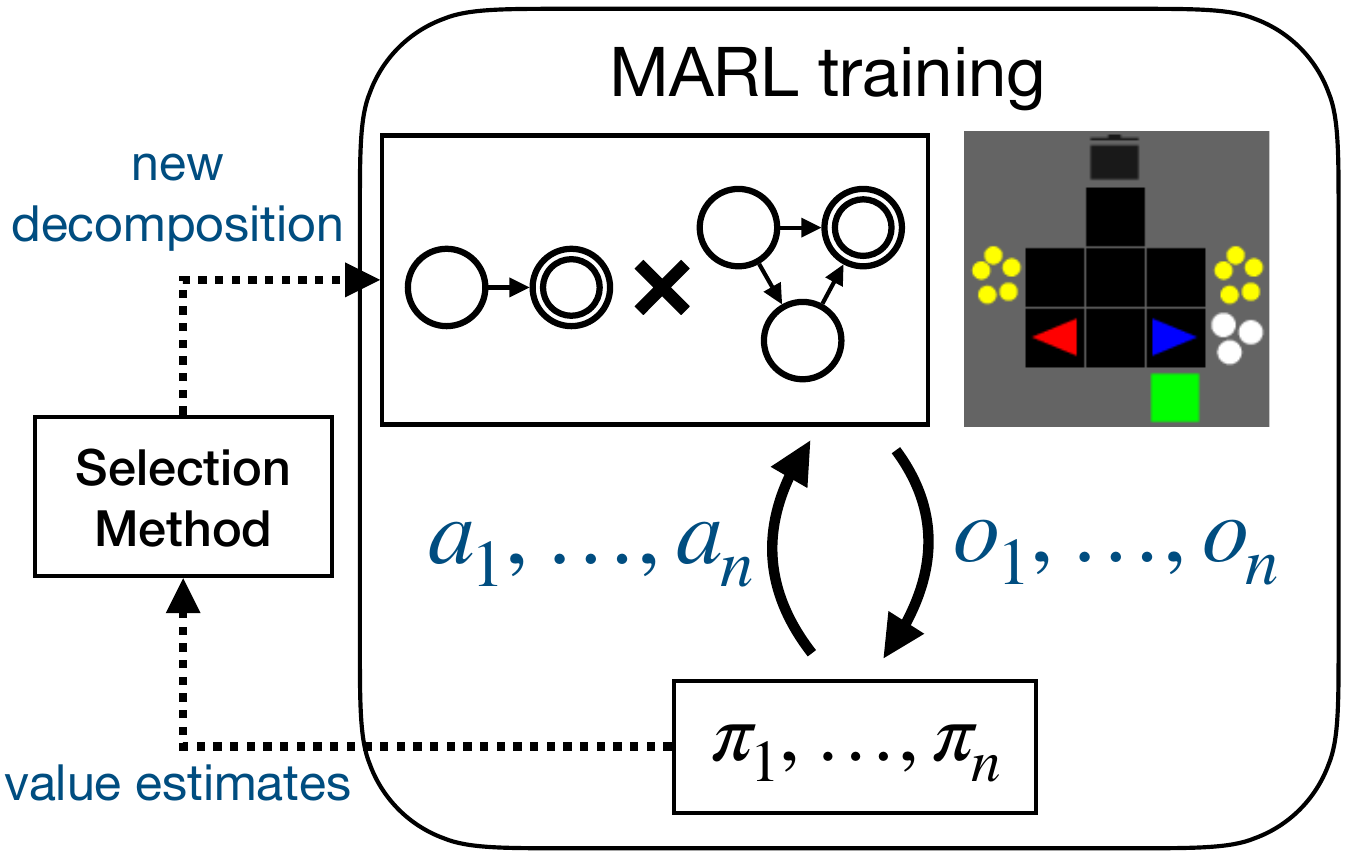}
\caption{Visualization of our learning framework. At each new episode of training, a selection method chooses a possible symbolic decomposition of the task and assigns sub-tasks to each agent in the team. As each agent learns the viability of different sub-tasks, our selection method simultaneously finds the optimal task decomposition.}
\Description{Teaser image.}
\label{fig:overview}
\end{figure}

These prior approaches to reward machine decomposition are limited when many possible decompositions exist. In these cases, the optimal decomposition of a task must be carefully designed and selected by a human. Selecting a meaningful task decomposition amongst many possible options often requires prior knowledge of the environment that is not assumed in standard RL, making the process tedious at best and impossible at worst. If a task decomposition is selected arbitrarily, the decomposition may not be compatible with the specifics of the environment to effectively break down the task. 

\begin{figure}[t]
\centering
    \begin{minipage}[b]{0.49\textwidth}
        \centering
        \includegraphics[width=0.85\linewidth]{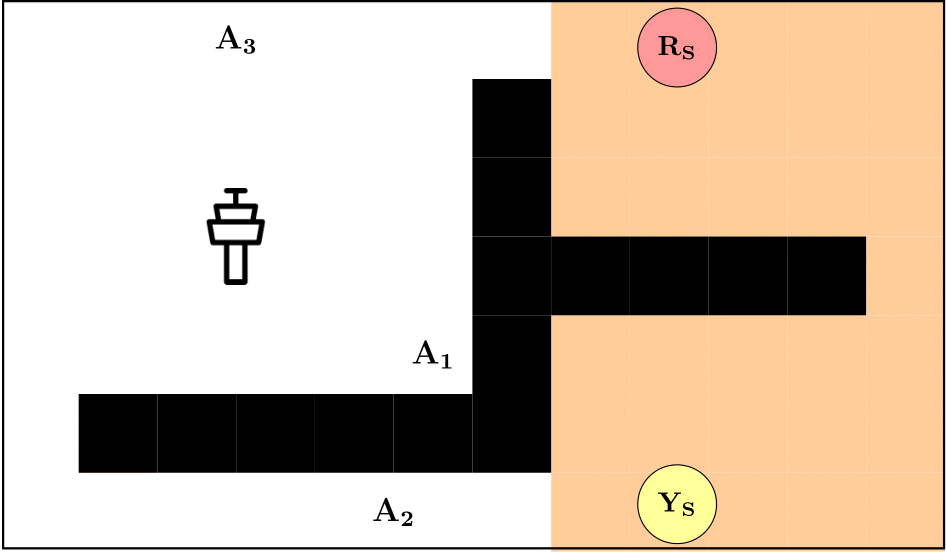}
    \end{minipage}
    \begin{minipage}[b]{0.49\textwidth}
    \includegraphics[width=0.99\linewidth]{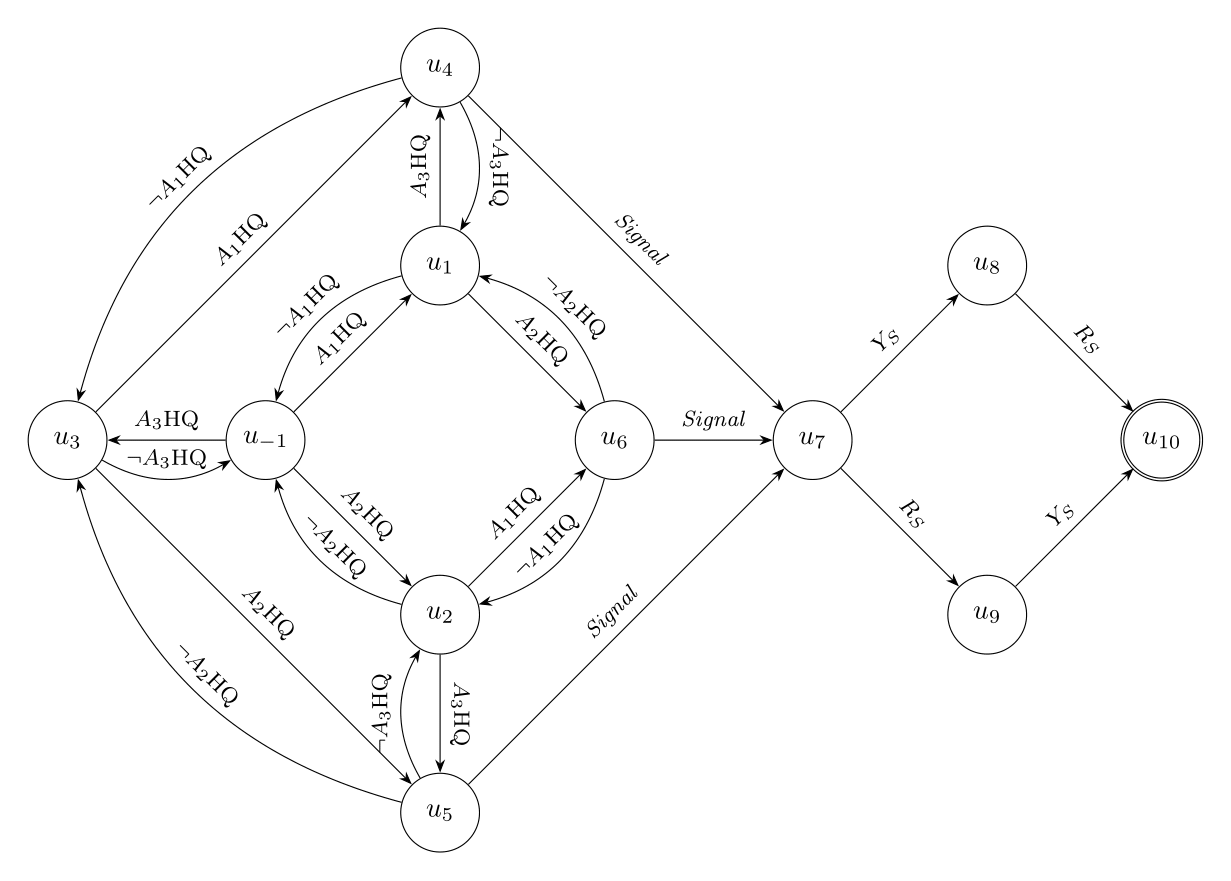}
    \end{minipage}
\caption{(Top) The ``Repairs'' MDP with a team of 3 agents. (Bottom) A task completion reward machine (RM) encoding the task: agents must navigate the environment to visit the HQ control tower, and then visit a set of communication stations. The goal state of the RM is denoted by concentric circles.}
\label{fig:running_example}
\Description{Visualization of our running example.}
\end{figure}

\textbf{Motivating example.} Consider the ``Repairs'' environment in Figure~\ref{fig:running_example}, where a team of three robot agents must visit a number of communication stations in their environment to make repairs. First, any two of the agents must meet at headquarters (HQ, denoted by the control tower symbol), at which point a ready signal is sent out to both stations (red and yellow) to inform the stations that agents will be visiting each station to perform repairs. The agents are then tasked to visit these stations in any order. The agents can independently move around in grid in any of the four cardinal directions, or remain still.  We can formulate this task in the form of a \textit{Reward Machine} (RM), an automaton that encodes the objective over high level `events' which occur in specific states of the environment (visualized in Figure~\ref{fig:running_example}). RMs offer a precise notion of task completion and can easily capture temporal dependencies required in objectives (i.e., visit the stations \textit{after} HQ).

In the Repairs environment, there exists a hazardous region, encoded in orange, that prevents more than one agent from entering the region at a time. 
The location and existence of this hazardous region is \textit{not} known to the agents a priori, which prevents this constraint from being encoded in the reward machine as part of the task description.

This task, even with no prior knowledge of the environment, can naturally be decomposed in a number of ways. For example, Agent 1 and Agent 2 can be tasked with meeting at HQ, and then Agent 2 can visit the yellow station and leave the hazardous region while Agent 3 visits the red station. Alternatively, Agents 2 and 3 can meet at HQ, and then Agent 1 can visit both the red and yellow stations. Although many plausible decompositions exist, knowing which decomposition of the overall task leads to the most efficient completion of the task (i.e., is \textit{optimal}) largely depends on the dynamics of the underlying environment. In our case, Agent 3 happens to be closest to the red station, which might mean that assigning that station to them would be optimal. However, without knowledge of the layout of the environment, as is standard in a model-free setting, this would be impossible to know ahead of time.


\textbf{Our Contributions.} In this work, we address the aforementioned limitations by introducing {\em an approach to automatically find optimal decompositions of an overall task into sub-tasks} for a team of agents. Our method is a lightweight extension of standard MARL and is applicable to model-free settings with no prior information about the environment or individual capabilities of the agents. 

We summarize our approach briefly as follows. At the beginning of training, we generate possible \textit{candidate} decompositions of our overall task as assignments of sub-tasks for our agents to accomplish. Then, during training, we explore selecting different decompositions for our team of agents and observe their performance as they attempt to learn a goal-conditioned policy that can achieve the variety of possible sub-tasks captured by our candidates. We record the value of agents' performances for different sub-tasks, and use this information to intelligently select subsequent decompositions for our agent team. This allows us to simultaneously learn (1) the optimal decomposition for our task and (2) policies for each agent that optimize said decomposition. We leverage selection strategies popularized in the multi-armed bandit algorithm literature \cite{katehakis1987multi, auer2002finite, audibert2009exploration} to balance exploring different candidate decompositions with exploiting the value our agents achieve as they learn sub-tasks during training. We visualize our approach in Figure~\ref{fig:overview}.

In addition to our decomposition selection strategy, we introduce a novel training setup for RM–driven MARL that rectifies issues caused by dependent agent dynamics. Prior approaches \cite{neary2020reward, smith2023automatic} to RM-driven MARL make the assumption that the dynamics of each agent are independent of other agents, so that each agent can be trained individually in an environment absent from other agents. This assumption is often impractical: it is inefficient to train individual agents independently, and in many cases, multi-agent dynamics are codependent. For example, an agent that has completed their sub-task may obstruct another agent from the completion of their own sub-task. Our setup gives each agent in the team a \textit{global} view of the overall task along with their sub-task, enabling agents to learn policies that accomplish their individual sub-tasks and help facilitate the completion of other agent's sub-tasks as well. 

We summarize our contributions as follows: \textbf{(1)} We introduce a method for learning optimal decompositions from model-free interactions with the environment. \textbf{(2)} Within our method, we provide a novel training architecture that allows multiple agents to train simultaneously within an environment and avoids conflicts that arise due to dependent dynamics across agents.
\textbf{(3)} We demonstrate our approach's improvement in learning multi-agent policies against baseline approaches in several cooperative environments. 
Our results show that MARL benefits substantially from the improved credit assignment of task decompositions even when the optimal decomposition is not known \textit{a priori}.


\section{Preliminaries}
\subsection{MDPs and Labeled Markov Games}
A Markov Decision Process (MDP) $\mdp = (S, A, T^\mdp, d_0, \gamma)$ is a tuple that consists of a state space $S$, an action space $A$, a transition function $T^\mdp: S \times A \rightarrow S$, an initial state distribution $d_0 \in \Delta(S)$, and a discount factor $0 < \gamma < 1$. A (stationary) policy $\pi: S \rightarrow \Delta(A)$ in $\mdp$ produces a distribution over actions given a state. Following standard RL notation, we define $\pi(a_t | s_t)$ as the probability of taking action $a_t$ in state $s_t$ at timestep $t$. A policy takes an action in a given state in $\mdp$, transitions to a new state via $T^\mdp$, and repeats, generating a \textbf{trajectory} of length $\tau$ as $(s_0, a_0, \dots s_\tau, a_\tau)$. 

To generalize to the MARL setting, we extend $\mdp$ to a cooperative Markov game with homogeneous action spaces. We define a cooperative Markov game with $n$ agents as $\markovgame = (\states, \actions, T^\markovgame, d_{0^1} \dots d_{0^n}, \gamma, L^\markovgame$), which corresponds to the joint set of states $\states = S_1 \times \dots \times S_n$, the joint set of actions $\actions = A_1 \times \dots \times A_n$, a joint transition operator $T^\markovgame: \states \times \actions \rightarrow \states$, a set of independent initial distributions for each agent $d_{0^1} \times \dots \times d_{0^n}$, and
a discount factor $\gamma$ that remains unchanged. We consider the \textit{centralized training, decentralized execution} \cite{bernstein2002complexity} setting of MARL, where each agent receives independent observations and deploys their individual policies $\pi_i: S_i \rightarrow \Delta(A_i)$ at execution time, but are jointly trained. The joint policy $\boldsymbol{\pi}: \states \rightarrow \Delta(\actions)$ executes all individual policies simultaneously at each timestep, and successive states are dictated by $T^\markovgame$, generating joint trajectories $((s_{0^0}, a_{0^0},\dots s_{0^n}, a_{0^n}), \dots (s_{\tau^0}, a_{\tau^0},\dots s_{\tau^n}, a_{\tau^n}))$. 

In addition to the standard components of our Markov Game, we assume access to a known \textit{labeling function} $L^\markovgame: \states \rightarrow 2^\Sigma$ that maps states in $S$ to a set of \textit{environment events}, denoted by $\Sigma$. Each environment event $e \in \Sigma$ is represented by a variable that takes on a Boolean truth value (True or False). For example, in Figure~\ref{fig:running_example}, if Agent 1 is in the yellow station, Agent 2 is in the red station, and Agent 3 has not moved from its initial position, the labeling function for this joint state would return $\{Y_S, R_S\}$.

\subsection{Task completion Reward Machines} 
In this work, we consider task completion reward machines (RMs)~\cite{xu2020joint, neary2020reward, smith2023automatic} as our team objective. A task completion RM is specified by a tuple $\rewmach = (\rmstates, u_{-1}, \Sigma, \delta, \rmstates^*)$ consisting of a set of states $\rmstates$, an alphabet of events $\Sigma$ that trigger transitions in $\rewmach$ via the transition function $\delta: \rmstates \times \Sigma \rightarrow \rmstates$, an initial state $u_{-1}$, and a set of goal states $\rmstates^*$. There are no outgoing transitions from any goal state $u^* \in \rmstates^*$. Task completion RMs additionally define an output scoring function $\sigma: \rmstates \times \rmstates \rightarrow \mathbb{R}$, where $\sigma(u, u') = 1$ whenever $u \notin \rmstates^*$ and $u' \in \rmstates^*$, and $0$ otherwise. Task completion RMs are Mealy machines \cite{mealy1955method} where reaching the goal state represents a valid completion of the task. We remark that $\delta$ and $\sigma$ are partial functions defined on subsets of $\rmstates \times \Sigma$ and $\rmstates \times \rmstates$.
The transition operator for a task completion RM takes in single events $e$ in as input; when multiple events occur simultaneously, the events are passed in sequence to the RM in arbitrary order. 

A task completion RM is connected to a Markov game by the labeling function $L^\markovgame$. We can project a trajectory in $\markovgame$ to a trajectory in $\rewmach$ by applying $L^\markovgame$ to each state, creating a sequence of event sets $(\{e_i \dots e_k\}_0, \dots \{e_i \dots e_k\}_\tau)$. This sequence of event sets transitions $\rewmach$ to create a trajectory $(u_0, \dots u_{\tau})$, where $u_0$ is the state resulting from $\delta(u_{-1}, \{e_i \dots e_k\}_0)$, and so forth. We say $\rewmach$ \textit{accepts} a trajectory if  $u_{\tau} \in \rmstates^*$. By definition, the cumulative score of $\sigma$ will be 1 for accepting trajectories, and 0 for all non-accepting trajectories.

\subsection{Using task completion RMs in MARL}

Recall our problem setting of \textit{centralized training, decentralized execution}: each agent will receive independent observations during execution. In other words, each agent will view the cooperative Markov game as an MDP, where the presence and actions of other agents are captured by the dynamics of their respective environments. 
To train our agent team, we can use the acceptance condition of a task completion RM as a reward function for MARL. This objective can be naturally compiled down to a reward function expressed over a \textbf{product MDP} for individual agent policies $\pi_1 \dots \pi_n$. Concretely, a product MDP synchronizes $\mdp$ and $\rewmach$ so that an agent may learn a policy over the joint space by coupling the reward machine state $u_t$ during a trajectory with the MDP's state $s_t$, producing an action conditioned on both: $\pi(a_t | s_t, u_t)$. If the agents transition to a goal state in $\rewmach$, they will receive a reward of one; all other transitions will receive a reward of zero. Existing deep RL algorithms, such as PPO~\cite{schulman2017ppo}, have shown success in learning performant policies for RMs in a variety of single-agent settings by learning policies over the product MDP~\cite{Li2024noisyuncertain, voloshin2023eventual}.

However, in a MARL setting, providing the full task completion RM for each agent creates difficulties in learning. If all agents see the same states and transitions in $\rewmach$, then all agents will receive credit when the task is completed, even if one or more agents did not contribute to completion of the task. Recall our running example in Figure~\ref{fig:running_example}. Suppose Agent 1 and Agent 2 visit HQ at the same time, and then Agent 2 visits the yellow and red stations, while Agent 3 remains stationary. Because the task is completed, all agents will receive equal reward for that episode. As a result, Agent 3 will think that its stationary behavior is desirable and be encouraged to act similarly in future episodes. 

Existing work addresses this shortcoming by introducing the notion of \textit{task decomposition} for a given task completion RM~\cite{neary2020reward}. A decomposition of a task completion RM creates sub-tasks in the form of $n$ smaller RMs $\rewmach_1 \dots \rewmach_n$, derived from the original, that can then be assigned to each agent. Each agent is only concerned with accomplishing their own sub-task encoded by the RM assigned to them. In order to compute a decomposition, a practitioner provides \textit{Local Event Sets} (LES) for each agent. An LES is a subset of events $\Sigma_i \in \Sigma$ that is deemed relevant to agent $i$'s individual sub-task and restricts agent $i$ to only observing events in $\Sigma^i$. A task completion RM is then \textit{projected} onto these local event sets to create an individual's sub-task reward machine $\rewmach_i = (\rmstates_i, u_{-1^i}, \Sigma_i, \delta_i, \rmstates^*_i)$. The states, initial state, and goal states of $\rewmach_i$ are sets of equivalence classes over states in $\rewmach$ based on an equivalence relation that subsumes states whose transitions do not contain any event in $\Sigma_i$. The transition function is a projection of the original $\delta$ where a transition between two states $u_j$ and $u_k$ in $\rewmach_i$ exists only if an event in $\Sigma_i$ triggered a transition between two distinct states in $\rewmach$ that were subsumed by $u_j$ and $u_k$ respectively. For the exact procedure of this projection, see~\cite{neary2020reward}. 

For each agent's sub-task RM $\rewmach_i$, we define an individual labeling function $L_i$ that projects the set of environment events returned by $L$ to the events belonging to an agent's local event set $\Sigma_i$. We say an event $e$ is a \textit{shared event} if it belongs to more than one local event set. For example, the event \textit{"Signal"} in Figure~\ref{fig:running_example} is a shared event. In the case of shared events, agents' sub-task RMs must \textit{synchronize} on this event. This means that a synchronized event must trigger a transition for all agents' RMs that share the event (i.e. all sub-task RMs must be in the appropriate state for the synchronized event to cause a transition), or no transition is taken from encountering that event for any agent's RM. 

The decomposition approach outlined in~\cite{neary2020reward} relies on a practitioner manually designing the local event sets to assign to each agent.  
In order to help guide the search for an appropriate assignment of local event sets,~\cite{neary2020reward} provides a notion of \textit{validity} for a given decomposition: a decomposition is valid if and only if the parallel composition of all reward machines $\rewmach_1 \dots \rewmach_n$ is bisimilar to the original $\rewmach$. If a decomposition is valid, then for any trajectory of events $\xi$, $\rewmach$ accepts $\xi$ if and only if all sub-task RMs $\rewmach_1 \dots \rewmach_n$ accept $\xi$.
In other words, a trajectory of events that accomplishes all decomposed sub-tasks encoded by $\rewmach_1, \dots, \rewmach_n$ is guaranteed to solve the overall task.

\section{Methodology}
In practice, multiple valid decompositions often exist for a given RM. However, a valid decomposition may not be \textit{feasible} under the dynamics of a given MDP; that is, the resulting learned policies may not be able to achieve the individual tasks prescribed by a decomposition. Moreover, when multiple feasible decompositions do exist, we do not know which decomposition most efficiently achieves the task or provides the best credit assignment for learning. Recall the Repairs environment and task in Figure~\ref{fig:running_example}. A feasible decomposition of the task would be for Agent 1 and Agent 2 to meet at HQ, then for Agent 2 to visit the yellow, then red stations in that order. However, this decomposition is less efficient than a decomposition where Agents 1 and 3 visit HQ, then Agent 2 visits the yellow station and leaves the hazardous region while Agent 3 visits the red station (assuming Agents complete their sub-tasks with optimal efficiency).

Recent work attempts to automate the search for RM decompositions by leveraging additional information provided by a practitioner~\cite{smith2023automatic}. In this work, subsets of events in $\Sigma$ can be provided that either require or forbid an event to belong to a specific agent's local event set $\Sigma_i$, along with a utility function that quantifies how valuable an event would be to a specific agent. This information is then used to find a subset of events in $\Sigma$ that still leads to a valid decomposition of $\rewmach$, if one exists. However, the aforementioned approach does not leverage knowledge gained about the MDP during training and therefore cannot ensure that the decomposition generated from their method is feasible or optimally efficient. In what follows, we will introduce our approach, which aims to find the optimal decomposition by learning a policy on-the-fly for many possible sub-tasks during training.

Our approach can be broken down into three primary components: \textbf{(1)} automatically generating a set of possible (candidate) decompositions for our task, \textbf{(2)} using a task-conditioned policy to generalize learning across multiple decompositions, and \textbf{(3)} employing the Upper Confidence Bounds (UCB) strategy~\cite{auer02ucb} to balance exploration and exploitation of candidates throughout training. 

\subsection{Generating candidate decompositions}
\label{subsec:generating_decompositions}
Approaches to procedurally generate decompositions of reward machines and similar automaton-based task specifications have been explored in the literature~\cite{smith2023automatic, lauffer2022dfadecomp}. Such methods can be used to generate a finite set of candidate decompositions $\decompositions = \{(\rewmach_{1}^1 \dots \rewmach_{n}^1), \dots \\(\rewmach_{1}^{|\decompositions|} \dots \rewmach_{n}^{|\decompositions|})\}$ from which the optimal decomposition can be found. We will denote an individual decomposition in our set as $d \in \decompositions$. In our work, we will use the decomposition approach for task completion RMs introduced in~\cite{smith2023automatic}. In this approach, all possible \textit{valid} decompositions are generated given $\Sigma$, and each one is assigned a score based on three factors: (1) minimizing the average number of events assigned to each agent's local event set, (2) maximizing the similarity amongst the sizes of all local event sets, and (3) maximizing the average `utility' of each agent's local event set, based on a practitioner-provided utility function that maps the assignment of an event to an agent to a scalar value. In our work, we assume that no utility function has been provided. We will therefore enumerate the `top-$k$' valid candidate decompositions based on a weighted sum of scores pertaining to factors (1) and (2), where $k$ is a hyperparameter that sets the number of decompositions we will consider, i.e. $k = |\decompositions|$. Our objective is to find the decomposition $d^* \in \decompositions$ that leads to a learned policy which will most efficiently achieve the original task $\rewmach$.

\begin{figure*}[t]
\centering
\includegraphics[width=0.9\linewidth]{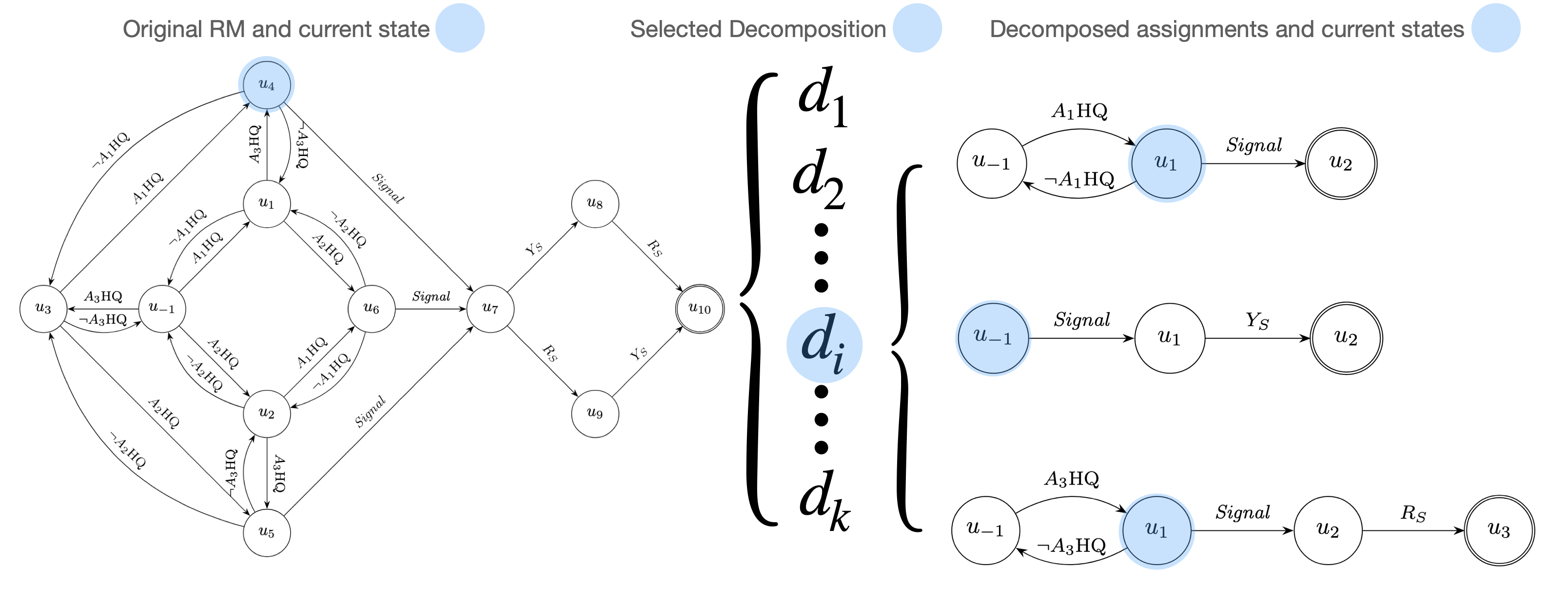}
\caption{A visualization of the information each agent receives using the policy architecture described in section~\ref{subsec:task_conditioned_setup} for the Repairs task from Figure~\ref{fig:running_example}. In addition to the observation gathered from the MDP, each agent's policy is conditioned on (1) the current state of the original RM task, (2) which decomposition is currently selected, and (3) the current state of their assigned sub-task RM within the selected decomposition.
}
\label{fig:observation_viz}
\Description{Visualization of the information provided to agent policies.}
\end{figure*}

\subsection{Task-conditioned policies}
\label{subsec:task_conditioned_setup}
We consider each agent policy in our team as sharing a \textit{task-conditioned} policy. Instead of learning separate policies for individual sub-tasks, we learn a single policy that outputs actions conditioned on a specific sub-task RM belonging to a decomposition generated in section~\ref{subsec:generating_decompositions}. We can then deploy separate copies of our policy on each agent during execution time.
\paragraph{Policy Architecture} Our policy is represented by a feedforward neural network that is shared and learned across all agents. The neural policy receives four inputs, depending on the individual agent:
\begin{itemize}
    \item an observation from the underlying MDP (different for each agent);
    \item an encoding of the selected decomposition $d_j \in \decompositions$ for the current execution episode (same for each agent); 
    \item the current state in their sub-task $\rewmach_{i}^{j} \in d_j$ (different for each agent);
    \item an encoding of the overall task and the current state in the overall task (same for each agent).
\end{itemize}

We visualize these inputs in Figure~\ref{fig:observation_viz} for an example decomposition of our Repairs task.
Learning a task-conditioned policy allows us to distinguish between different sub-tasks within different decompositions while exploiting the generalization capabilities of multi-task learning~\cite{qiu2023gcrlltl, vaezipoor2021ltl2action} as well as the natural curriculum learning inherent in decomposition exploration. Let us return to the running example from Figure \ref{fig:running_example}. Imagine that as part of a decomposition selected during training, agent $A_1$ is only assigned the task of meeting at HQ and nothing else. Even if this decomposition is not optimal, succeeding in this sub-task teaches $A_1$ the valuable skill of how to reach HQ. This experience will be useful in the future when $A_1$ is tasked with more complicated sub-tasks, such as \textit{``first go to HQ and then the red station"}.

Conditioning on the overall task allows us to relax the assumption of independent dynamics as we will describe in Section \ref{sec:indep_dyn}. In our experiments, we use one-hot encodings of the selected sub-task and the current position in the sub-task during the rollout. However, the embeddings could in principle be generated in any way, such as learned embeddings from a graph neural network that encode the structure of an RM~\cite{yalcinkaya2023automata}.

\subsection{Selecting decompositions during training}
\label{subsec:decomp_strategies}

A naive approach to our objective of finding an optimal ordered decomposition would be to learn a task-conditioned policy for each $d \in \decompositions$ independently, and then choose the decomposition that performs the best after a certain amount of training. As the number of candidate decomposition increases, this approach quickly becomes intractable. We seek to improve the efficiency of this process by simultaneously learning both the most efficient decomposition and the corresponding policies that achieve this decomposition. 

A key insight of our work is that we can use rewards from previous executions of a sub-task $\rewmach_{i}^{j}$ to estimate the satisfaction likelihood of that sub-task. These estimates, which we call \textit{value estimates}, are used as a heuristic in decomposition selection to assess how well a policy is performing on a specific sub-task. Our approach computes value estimates as an \textit{exponential weighted moving sum} of previous rewards, as more recent rewards are typically a more accurate reflection of the performance of the current policy.

We will denote the value estimates for a sub-task $\rewmach_{i}^{j}$ as $V_{\rewmach_{i}^{j}}$. On episode $H$ of training on decomposition $j$, we can compute a sub-task's value estimate as $V_{\rewmach_{i}^{j}} = \sum_{h=0}^H \alpha^{H-h} r_h$. Here, $ r_h$ represents the reward achieved by agent $i$ under sub-task $\rewmach_i$ from the $h$-th execution and $\alpha$ is a hyperparameter defining the decay rate.

With value estimates in hand for each agent policy in our team, on each sub-task, we can consider a number of heuristic approaches to utilize them. In this work, we use the Upper Confidence Bound (UCB) algorithm~\cite{auer02ucb}, which balances exploring different decompositions with exploiting higher scoring decomposition via a hyperparameter $\beta$. The score assigned to each decomposition $d_j$ is an average of the decomposition's current value estimates for each sub-task $\{V_{\rewmach_{1}^{j}}, \dots, V_{\rewmach_{n}^{j}} \}$.

We note that the value estimates early in training may be arbitrarily inaccurate due to the lack of progress made in learning the policy for different sub-tasks by each agent. When inaccuracy is high early in training, exploration is critical so that agents can sufficiently optimize towards achieving each candidate sub-task before the value estimates are exploited to converge on the optimal decomposition. Once the optimal decomposition is converged upon, additional training will further optimize the policy conditioned on its corresponding sub-tasks.

\subsection{Dealing with dependent dynamics} \label{sec:indep_dyn}
Prior work in RM-guided MARL \cite{neary2020reward, smith2023automatic} assumes that the underlying dynamics of the MDP are independent between agents. This prevents the use of an MDP with dynamics that model collisions or interactions between agents unless the interaction is explicitly modeled by the task's reward machine, which requires knowing about the interaction a priori. 

Consider the motivating example visualized in Figure~\ref{fig:running_example}. The example includes dependent dynamics between the agents in the form of the hazardous region. Only one agent can occupy the region at a time, meaning that an agent's ability to enter the region is dependent on the other agents' positions and actions.

Prior works \cite{neary2020reward, smith2023automatic} require the assumption of independent dynamics so that the completion of an individual agent's reward machine is independent of the behavior of other agents. With dependent dynamics, this is rarely the case. For example, imagine that Agent 3 is assigned to reach the yellow target and Agent 2 is assigned to reach the red target. If Agent 3 enters the hazardous region and reaches the yellow target, it accomplishes its sub-task. Now, completion of the overall task only requires Agent 2 to reach the red target. Since the red target is located in the hazardous region, this requires Agent 3 to exit the hazardous region. However, Agent 3 has already accomplished its sub-task (and has no knowledge of Agent 2's task), so it has no incentive to leave the hazardous region.

We are able to relax the assumption of independent dynamics made in previous work \cite{neary2020reward, smith2023automatic} and handle environments such as the one in Figure~\ref{fig:running_example} by allowing agents to condition not only on the status of their sub-task, but on the status of the overall task as well. In addition, agents are given a small reward for the completion of the overall task during training along with their primary reward for completing their assigned sub-task. This reward structure incentivizes agents to condition on the status of other agent's tasks (e.g., that Agent 2 needs to go to red) so that the overall task can be completed (e.g., Agent 3 is incentivized to leave the hazardous region).

\begin{figure*}[t]
    \centering
    {
        \includegraphics[width=0.32\textwidth]{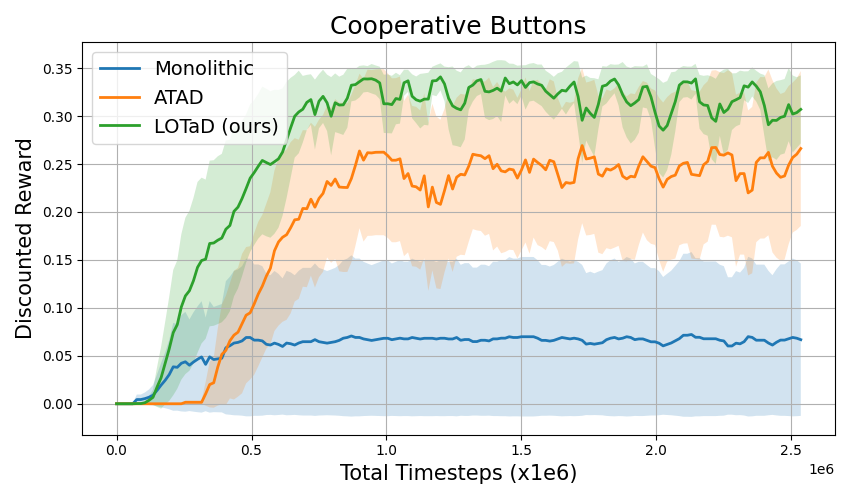}
        \label{fig:cyrus_results}
    }
    {
        \includegraphics[width=0.32\textwidth]{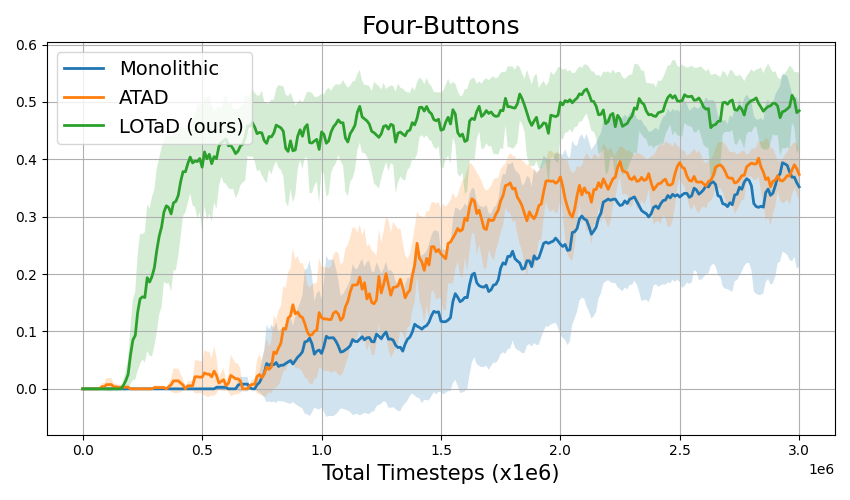}
        \label{fig:fig:challenge_results}
    }
    {
        \includegraphics[width=0.32\textwidth]{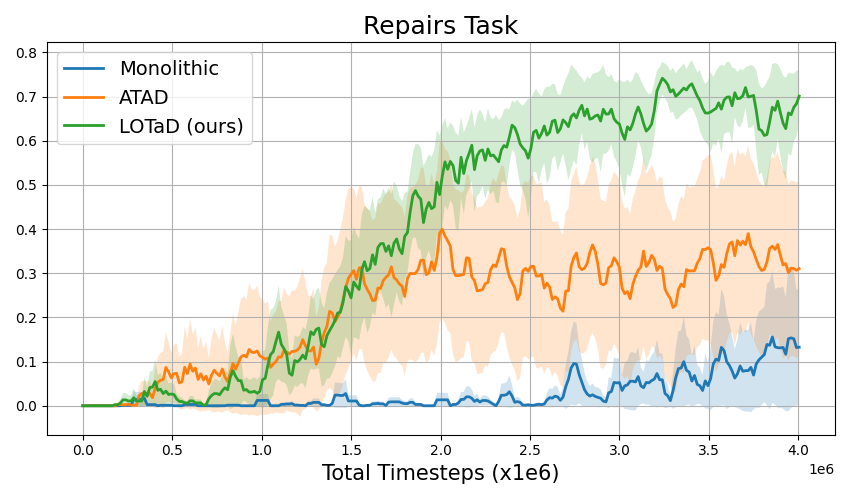}
        \label{fig:motivating_results}
    }
    {
        \includegraphics[width=0.32\textwidth]{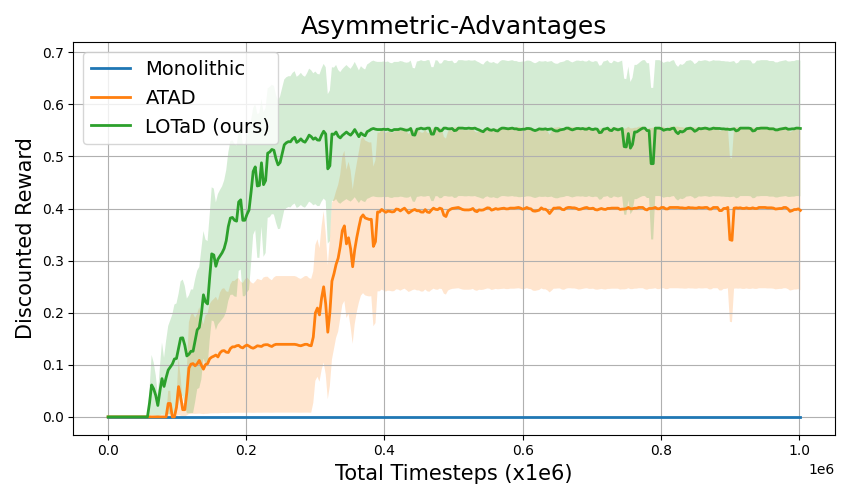}
        \label{fig:asymm_results}
    }
    {
        \includegraphics[width=0.32\textwidth]{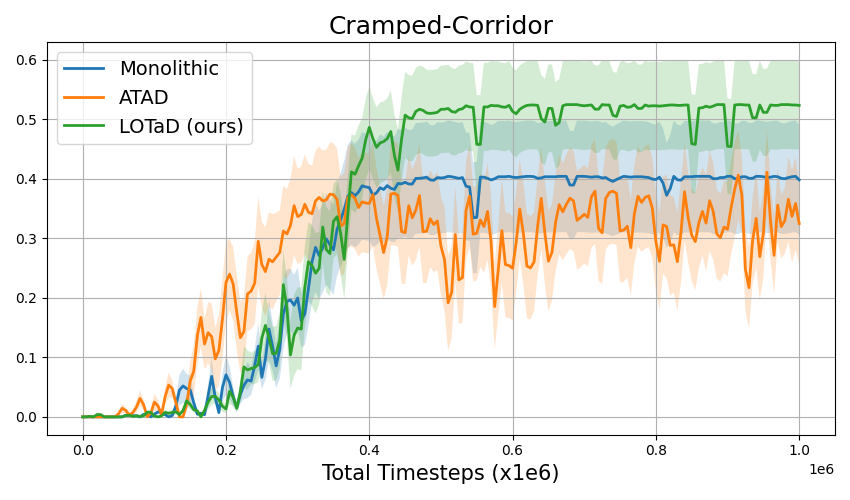}
        \label{fig:corridor_results}
    }
    \caption{Training curves for LOTaD and baseline methods in our experimental domains. Results are averaged over 5 random seeds.}
    \label{fig:combined_results}
\Description{Training curves for LOTaD and comparative baselines in our experimental domains.}
\end{figure*}

\section{Experiments}
We evaluate our proposed approach, which we refer to as \textbf{LOTaD} (\textbf{L}earning \textbf{O}ptimal \textbf{Ta}sk \textbf{D}ecompositions), in a variety of MARL settings with varying task complexity. Our code is available at \href{https://github.com/thomasychen/LOTaD}{https://github.com/thomasychen/LOTaD}.

In our experiments, we seek to answer the following research questions: \textbf{(RQ1)} Does LOTaD outperform existing approaches to RM-guided MARL that are not learning-informed, including approaches that do not perform task decomposition? \textbf{(RQ2)} Can LOTaD learn to avoid pitfalls in learning that may occur due to dependent dynamics across agents? \textbf{(RQ3)} How does varying the number of candidate decompositions $k = |\decompositions|$ affect the learning performance of LOTaD?

\subsection{Environments and tasks}

Our environments include the \textbf{Repairs} environment and RM task presented in Figure~\ref{fig:running_example}. The environment is instantiated as a 7x12 grid in which agents can move in any of the cardinal directions, or take a no-op action where no movement occurs. We make the Repairs environment stochastic by giving a small chance of an agent ``slipping'' and taking a random action rather than the action selected by their policy. Each agent observes only their position in the world at each timestep.

In addition to the Repairs environment, we also include two ``Buttons'' environments that require a team of agents to press a series of buttons in a particular order. We instantiate two environment-task pairs: First, we use \textbf{Cooperative Buttons}, a task from~\cite{neary2020reward} where any agent must reach a specified goal location, but in order to do so, must traverse regions that can only be crossed once a corresponding button has been pressed. Second, we use \textbf{Four-Buttons}, a task where two agents must press four buttons (yellow, green, blue, and red) in an environment, with an ordering constraint that the yellow button must be pressed before the red button. Both of these environments are represented as 10x10 grids with the same observation space, action space, and ``slipping'' dynamics as the Repair environment. Visualizations of these environments are presented in our Appendix.

Lastly, we evaluate LOTaD on two environments from the popular multi-agent benchmark \textit{Overcooked}~\cite{carroll2019overcooked}. In both environments, we provide the same simple task of delivering a soup to the delivery station, which requires putting three onions in a pot, plating, and delivering the soup. Agents must coordinate depending on the dynamics of their environment to efficiently cook and deliver the soup. We use a \textbf{Cramped-Corridor} environment, where two agents must navigate around one another to reach the pot at the end of a small corridor in a cramped room, and an \textbf{Asymmetric-Advantages} environment, where two agents are in separate rooms, but have access to an asymmetric set of resources in their respective rooms. We visualize both Overcooked environments in our Appendix. The observation space and action space for these environments are the same as those provided in the Overcooked implementation from~\cite{flair2023jaxmarl} with the adjustment that we do not expose the number of onions in the pot or whether the soup has finished cooking in order to ensure that no information provided by the labeling function is redundant in the agents' observations.

 In all environments, we apply a discount factor $\gamma < 1.0$ to the reward offered by our RM to incentivize our agents to accomplish the task as efficiently as possible. We generate $k=10$ decomposition candidates in all experiments using the generation method described in Section~\ref{subsec:generating_decompositions} for LOTaD to search amongst. We provide additional information regarding each environment and task in our Appendix. 

\subsection{Baselines}

To evaluate LOTaD, we compare against a baseline that selects a decomposition using the \textbf{ATAD} method~\cite{smith2023automatic}. This approach selects a set decomposition prior to learning based on the scoring method described in Section~\ref{subsec:generating_decompositions}. We break ties between top scoring decompositions arbitrarily. In addition to this baseline, we also compare against a baseline approach that assigns each agent the overall task. We call this baseline \textbf{Monolithic} and use it to evaluate how MARL would perform if no decomposition of the RM was used. 
We use PPO~\cite{schulman2017ppo} with a Gaussian policy over the action space for each agent as our RL algorithm in every environment. 

\subsection{Results}

\begin{figure*}[t]
    \centering
    \begin{minipage}{0.48\textwidth}
        \centering
        \includegraphics[width=\linewidth]{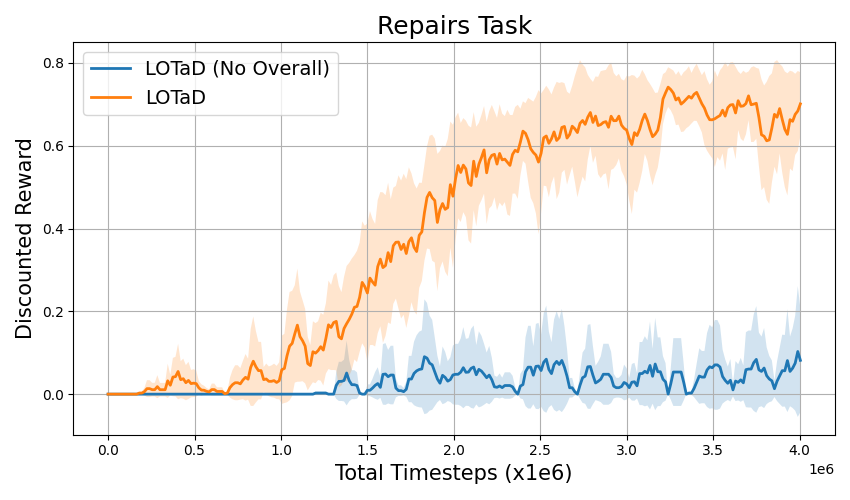}
    \end{minipage}
    \hfill
    \begin{minipage}{0.48\textwidth}
        \centering
        \includegraphics[width=\linewidth]{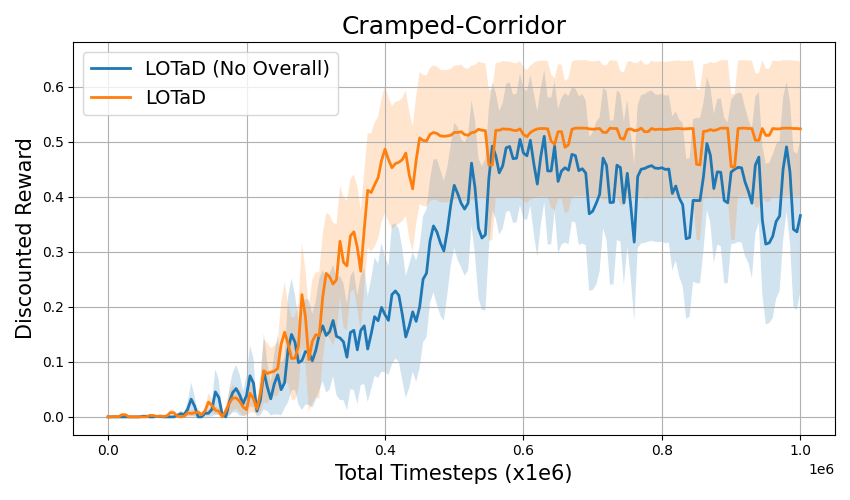}
    \end{minipage}
    \caption{Training curves for LOTaD in the Repairs Task and Cramped-Corridor environments demonstrating the effect of conditioning on the overall task state along with individual sub-task states for each agent.}
    \label{fig:no_mono_ablation_curves}
\Description{Training curves for our overall encoding ablation study.}
\end{figure*}

We plot training curves for all experimental domains in Figure~\ref{fig:combined_results}. In these curves, we report the current best discounted reward achieved by LOTaD amongst all decomposition candidates under consideration. We find that LOTaD outperforms both the monolithic and ATAD baselines across all environments, answering \textbf{(RQ1)} in the affirmative. In many of our environments, learning decompositions with LOTaD allows for agents to explicitly parallelize their contributions towards achieving the task. For example, in the Four-Buttons environment, a candidate decomposition allows for one agent to visit the yellow and green buttons while the other agent visits the blue button. 

Interestingly, LOTaD-learned decompositions are useful even when explicit parallelization is not possible, such as the Overcooked RM task. Decompositions of this task can still facilitate multi-agent learning: for example, one agent may be tasked with putting all three onions in the pot, while the other agent must wait for the soup to be cooked before plating and delivering the soup. In this decomposition, the agent tasked with plating and delivering the soup may fetch a plate and stand near the pot so that the soup can be quickly delivered upon completion of cooking.

In comparison to LOTaD, the ATAD baseline is unable to consistently find performant task decompositions. Although ATAD selects a decomposition based on the same scoring method by which we select $\decompositions$, there exist many possible decompositions tied for the highest achievable score in a given task. As a result, ATAD may select the optimal decomposition when the optimal decomposition is also the highest scoring, but performs suboptimally when this is not the case. In addition to lower reward, this leads to a higher variance in performance by ATAD, as evidenced by Figure~\ref{fig:combined_results}. The Monolithic baseline consistently achieves the lowest reward due to the sparsity of the overall task reward. Similar to ATAD, we notice that policy learning with the Monolithic baseline is unreliable and tends to converge more slowly than LOTaD.

We find that LOTaD is able to successfully accomplish the task in the Repairs and Overcooked Cramped-Corridor environments, which both involve dependent dynamics amongst agents. LOTaD is indeed able to avoid issues when training agent teams in environments with dependent dynamics, affirming \textbf{(RQ2)}. To further investigate our hypothesis of whether a global task view alleviates these issues, we ran LOTaD \textit{without} an encoding of the overall task in the Repairs and Cramped-Corridor environment. We present the training curves from this ablation study in Figure~\ref{fig:no_mono_ablation_curves}. We find that without the overall task encoding, LOTaD is largely unable to accomplish the Repairs task, and is slower and less stable in accomplishing the Cramped-Corridor task. This suggests that the overall task encoding enables our agents to learn policies that progress towards completion of the overall task even when an agent's individual sub-task is already achieved. 
In our other environment settings, where codependent agent dynamics do not exist, removing the overall task encoding does not significantly affect training.

\begin{figure}
\centering
\includegraphics[width=0.9\columnwidth]{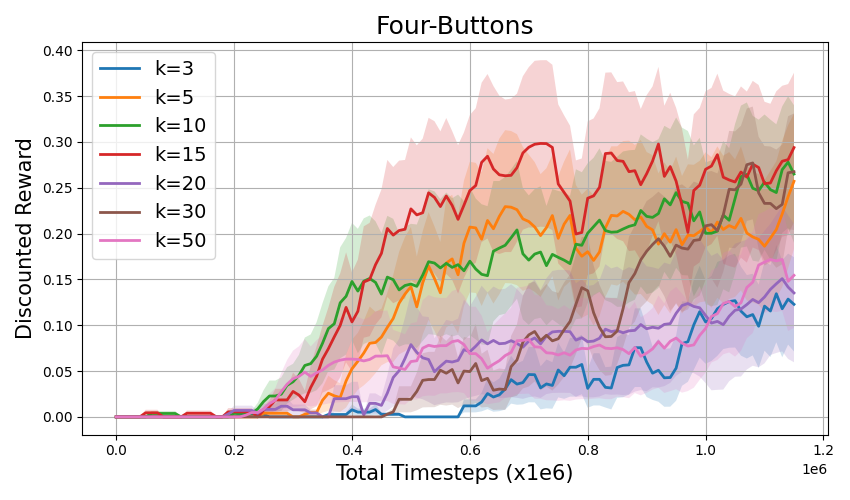}
\caption{Training curves for the UCB selection strategy on increasingly sized $\decompositions$ for the Four-Buttons task. Results are averaged over 10 random seeds.}
\Description{Ablation study varying LOTaD's performance on varying decomposition set sizes.}
\label{fig:scaling_results}
\end{figure}

To answer \textbf{(RQ3)}, we perform ablation studies where we vary the size of $\decompositions$. These training curves are visualized in Figures~\ref{fig:scaling_results}. We see that values of $k$ that are either very small or very large make the learning problem more challenging for LOTaD. We reason that if $k$ is too large, more exploration is required, and LOTaD may struggle to find the optimal decomposition amongst many candidates. If $k$ is too small, there is a lower chance of ``good'' decompositions appearing in the set $\decompositions$, which inherently limits the performance potential of LOTaD. We note that the inherent randomness of LOTaD, as well as the choice of environment, task, and the amount of exploration prioritized by our UCB hyperparameter $\beta$, may confound results.

\section{Related Work}
\paragraph{Automaton-based task specifications in RL} 
An extensive body of work has explored automaton-based task specification for RL agents. Previous efforts have proposed RL approaches to policy learning for reward machines~\cite{icarte2022reward, Icarte2020RewardMachine, Camacho2019LTLAndBeyond} or automata-based representations of temporal logic~\cite{VoloshinLCP2022, voloshin2023eventual, hasanbeig2020deep, alur22ltlframework, sadigh2014learning, FuLTLPAC, jothimurugan2019composable, shah2024ltldeeprl} by augmenting the state space of the MDP. These efforts focus primarily on single-agent settings and are not designed to handle the MARL case with shared objectives or formally decompose the automaton.

\paragraph{(Symbolic) task decomposition for multi-agent teams}
Symbolic structures have been leveraged to facilitate efficient multi-agent learning in a variety of settings. A number of these approaches rely on a known dynamics model~\cite{karimadini2011cooperative, schillinger2016fltldecomposiiton, schillinger2018simultaneous} for planning-based approaches. In contrast, we assume no prior knowledge of the environment dynamics as is standard in reinforcement learning.

Closely related to our contributions, previous works explored learning decompositions of tasks for teams of agents. Some previous works explored decomposing symbolic tasks by hand designing decompositions or using task- and environment-agnostic heuristics \cite{neary2020reward, smith2023automatic}. However, these works either require extensive human involvement in determining the optimal decomposition or do not offer a way to choose a decomposition based on MDP dynamics. Moreover, these approaches are limited to MDPs in which the dynamics of agents are independent. Other works learn role assignments for traditional reward functions (i.e., non-symbolic) with multi-agent teams \cite{wang2020rode} or decompose traditional (non-Markovian) reward functions \cite{sun2020reinforcement} for teams of agents but these methods do not easily extend to non-Markovian rewards which we consider.

\paragraph{Credit assignment in multi-agent RL}
Credit assignment, originally explored in the single-agent setting \cite{sutton1984temporal, riedmiller2018learning, ferret2019self, van2021expected}, is a problem that has been extensively explored in multi-agent learning literature and can be divided between \textit{explicit} and \textit{implicit} solutions. Explicit credit assignment most closely resembles our work, typically assuming that value functions are given in a specific form that allows certain types of decompositions, such as additive decompositions \cite{nguyen2017policy}, or 
value factorization \cite{wang2021towards, sunehag2017value}. Other explicit methods are based on assuming a hierarchical execution structure \cite{agogino2004unifying, feng2022multi, rashid2020monotonic}.
Implicit credit assignment instead attempts to perform credit assignment without an explicit structure, for example, by deriving individual policy gradients for each agent derived from a centralized critic \cite{zhou2020learning}. 

\section{Conclusion}
We introduce a novel approach for learning the optimal decomposition of a task completion RM for MARL through model free interactions with the environment. Our framework 
simultaneously learns the optimal decomposition and the policies that solve that decomposition amongst a set of candidates. Our experiments show that we improve the sample efficiency of multi-agent learning in model-free deep RL settings even when the optimal decomposition is not known a priori. 
Through our ablations, we have shown that both including the encoding of the overall task and intelligent decomposition selection is critical for sample efficient learning of reward machines in multiagent settings. 

We see a number of compelling directions for future work. For example, we are interested in exploring how different representations of an RM may enable greater sample efficiency by exploiting semantic similarity amongst overlapping sub-tasks. 
Lastly, we are interested in exploiting the inherent curriculum present in decomposition exploration by generalizing our multi-agent setting to solving multiple tasks, building on prior work in goal-conditioned RL on automata-based tasks~\cite{vaezipoor2021ltl2action, qiu2023gcrlltl}.

\begin{acks}
This work was supported in part by DARPA Contract FA8750-23-C-0080 (ANSR) and
the DARPA Contract HR00112490425 (TIAMAT), by Nissan and Toyota under the iCyPhy center, 
and an NSF Graduate Fellowship.

\end{acks}



\newpage
\bibliographystyle{ACM-Reference-Format} 
\bibliography{logicrl, new}

\appendix
\section{Limitations}\label{appendix:limitations}
Our proposed approach, LOTaD, searches through a pre-defined set of decompositions $\decompositions$ to efficiently find a decomposition that optimally solves the task. However, the quality of our found decomposition will only be as good as the quality of $\decompositions$. If all candidates in $\decompositions$ are highly suboptimal, our approach will be limited in its efficacy. To avoid this, we use on a decomposition generation technique that relies on heuristics to create balanced decompositions amongst agents. However, there is no way to ensure a priori that these decompositions will be successful in the context of the unknown MDP dynamics. Studying how the generation of $\decompositions$ affects the learning outcome of LOTaD is left as a compelling direction for future work.

One of LOTaD's key features is that it provides a \textit{global} view of the overall task to each agent, so that agents can coordinate within the MDP to solve the overall task, even after having solved their respective sub-tasks. Providing this global view (in the form of the overall task RM) requires agents to receive the same information from a labeling function when environment events occur. This level of information sharing may be impractical in some cases. However, we reiterate that this does not mean that control is shared across agents. 



\section{Additional experimental details}
\label{appendix:experiment_details}

\begin{figure*}[t]
    \centering
    \begin{minipage}{0.245\textwidth}
        \centering
        \includegraphics[width=\linewidth]{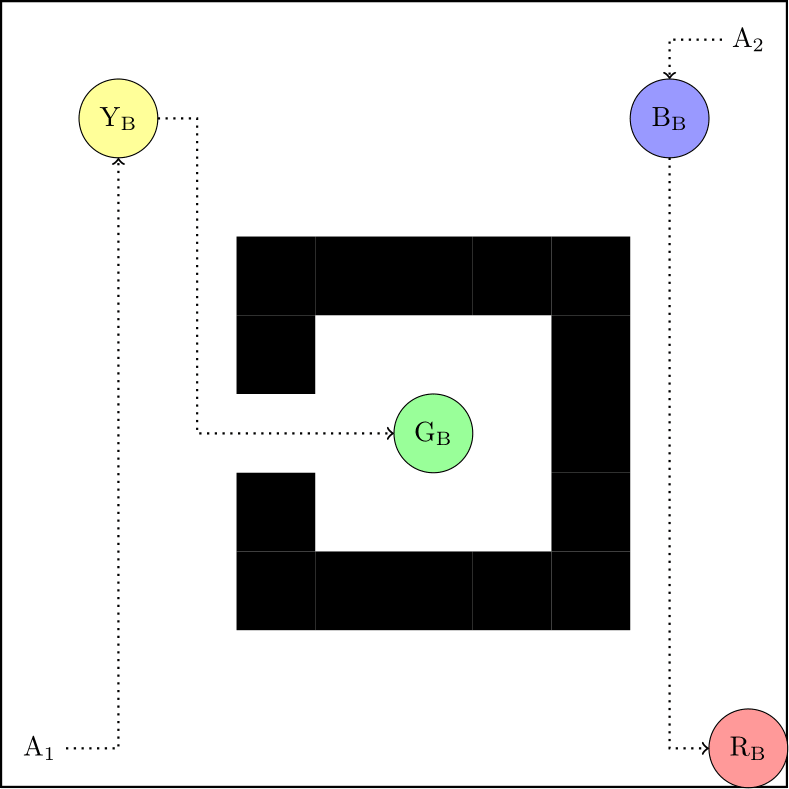}
    \end{minipage}
    \hfill
    \begin{minipage}{0.245\textwidth}
        \centering
        \includegraphics[width=\linewidth]{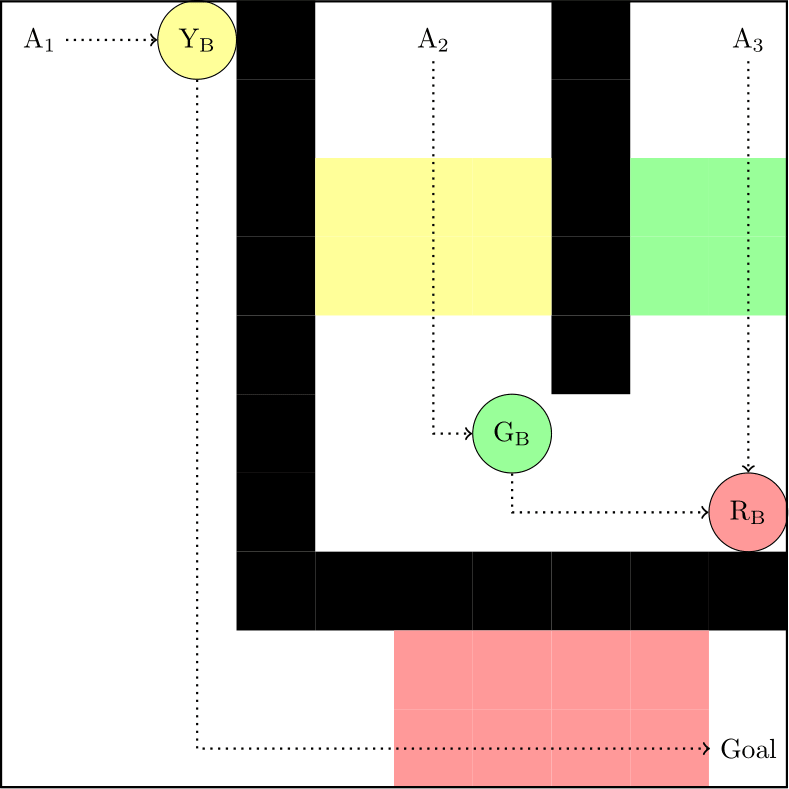}
    \end{minipage}
    \hfill
    \begin{minipage}{0.298\textwidth}
        \centering
        \includegraphics[width=\linewidth]{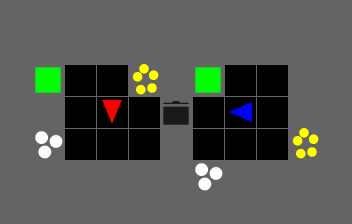}
    \end{minipage}
    \hfill
    \begin{minipage}{0.19\textwidth}
        \centering
        \includegraphics[width=\linewidth]{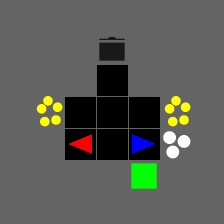}
    \end{minipage}
    \caption{From left to right: The Four-Buttons environment, The Cooperative Buttons environment, the Asymmetric-Advantages Overcooked Environment, and the Cramped-Corridor Overcooked Environment.}
    \label{fig:all_other_envs}
\Description{Environment visualizations.}
\end{figure*}

\subsection{Reward Machines}
In Table~\ref{tab:reward_machines}, we provide the task completion RMs used in each experimental domain. For the Overcooked domains (Cramped-Corridor and Asymmetric-Advantages), the same RM task is used. We omit sink state transitions from the Four-Buttons RM for the sake of visual clarity. These transitions would be taken if the event $R_B$ is triggered at states $\{u_{-1}, u_1, u_2, u_3, u_5, u_6, u_7, u_{10}\}$ and would lead to a sink (rejecting) state with no outgoing transitions.

As discussed in section~\ref{subsec:generating_decompositions}, we use the ATAD method from~\cite{smith2023automatic} to generate candidate decompositions. The ATAD method uses weight hyperparameters to balance the importance of condition (1) (decomposition size) and (2) (decomposition balance); we use weights 2 and 0.5 respectively in all experiments for both LOTaD and the ATAD baseline.
The sub-task RMs generated by this approach are called \textit{accident avoidance} RMs, and transition sub-task RMs to sink states if any event is experienced that does not belong to any local event set. We use these accident avoidance RMs as our sub-tasks, with the modification that and transitions to sink states receive a reward of $0$ rather than a reward of $-1$.

 ATAD also allows for the provision of forbidden and required event sets, which specify certain environment events that either cannot or must appear in a specific agent's local event set, respectively. The forbidden event sets for the Repairs environment were $A_1 = \{A_2\text{HQ}, \neg A_2\text{HQ},A_3\text{HQ}, \neg A_3\text{HQ}\}$, $A_2 = \{A_1\text{HQ}, \neg A_1\text{HQ},A_3\text{HQ}, \neg A_3\text{HQ}\}$, and $A_3 = \{A_2\text{HQ}, \neg A_2\text{HQ}, A_1\text{HQ}, \neg A_1\text{HQ}\}$. These forbidden event sets prohibit events that are specific to other agents from appearing in an agent's local event set. The required event sets for the Repairs environment are $\{ \text{Signal}\}$ for all agents. For the Cooperative Buttons task, the forbidden event sets are $A_1 = \{A_{2}^{R_B}, A_{2}^{\neg R_B}, A_{3}^{R_B}, A_{3}^{\neg R_B}\}$, $A_2 = \{A_{3}^{R_B}, A_{3}^{\neg R_B}\}$, and $A_3 = \{A_{2}^{R_B}, A_{2}^{\neg R_B}\}$, following a similar pattern to the Repairs environment. For the Four-Buttons task and the Overcooked task, both the forbidden and required event sets for all agents are empty. All decompositions generated by ATAD are required to comply with these forbidden and required event sets. We note that the provided forbidden and required event sets do not exploit any information about the environment and are independent of the MDP's dynamics.

 \begin{table*}[ht]
    \centering
    \begin{tabular}{lccccc}
        \hline
        \textbf{Hyperparameter} & \textbf{Four-Buttons} &  \textbf{Cooperative Buttons} & \textbf{Repairs Task} & \textbf{Cramped-Corridor} & \textbf{Asymmetric-Advantages} \\
        \hline
        Learning rate          & 5e-4   &  5e-4     & 5e-4   &2.5e-4  &2.5e-4\\
        Batch size             & 256      &  256        & 256    &256  & 256\\
        Number of epochs       & 5        &  5          & 5  & 5 & 5\\
        Discount               & 0.97      &  0.95       & 0.95  & 0.99 & 0.99\\
        Entropy Coefficient    & 0.1     &  0.1       & 0.1  & 0.01 & 0.01\\
        Max. grad. norm.       & 0.5      & 0.5         & 10  & 0.5 & 0.5\\
        Value loss coef.       & 0.5      & 0.5         & 10  & 0.5 & 0.5\\
        \hline
    \end{tabular}
    \caption{PPO training hyperparameters.}
    \label{tab:ppo_hyperparameters}
\end{table*}

\begin{table*}[ht]
    \centering
    \begin{tabular}{lccccc}
        \hline
        \textbf{Hyperparameter} & \textbf{Four-Buttons} &  \textbf{Cooperative Buttons} & \textbf{Repairs Task} & \textbf{Cramped-Corridor} & \textbf{Asymmetric-Advantages} \\
        \hline
        Exponential Decay $\alpha$        & 1   &  1     & 1   & 1  & 1\\
        UCB Exploration $\beta$      & 0.5   &  0.5   & 0.5  & 0.5  & 0.5\\
        Num. Candidates $k$         & 10   &  10  & 10 & 10  & 10 \\
        \hline
        Maximum Episode Length      & 100   &  100  & 400 & 400  & 200 \\
        \hline
    \end{tabular}
    \caption{Additional experiment hyperparameters.}
    \label{tab:hyperparameters}
\end{table*}

\label{appendix:reward_machines}
\begin{table*}[ht]
    \centering
    \begin{tabular}{|c|c|}
        \hline
        \textbf{Environment} & \textbf{Reward Machine Task}\\ 
        \hline
        Four-Buttons & \makecell{\centering \includegraphics[width=0.5\textwidth]{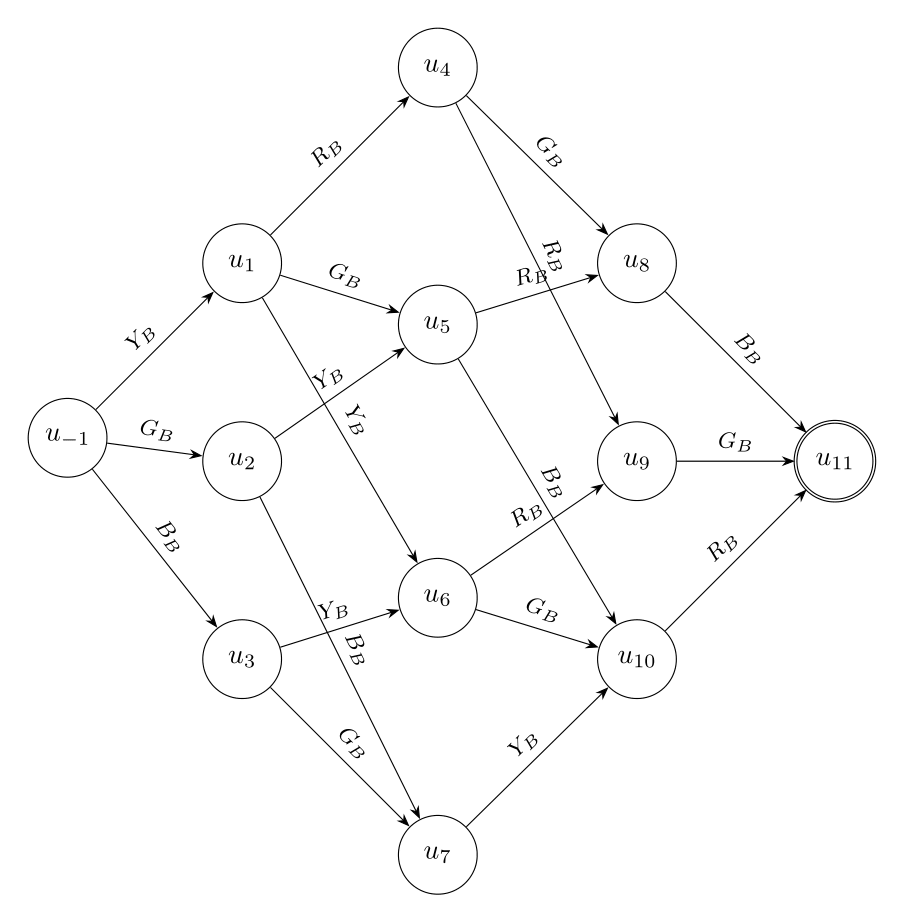}} \\
        Cooperative Buttons & \makecell{\centering \includegraphics[width=0.5\textwidth]{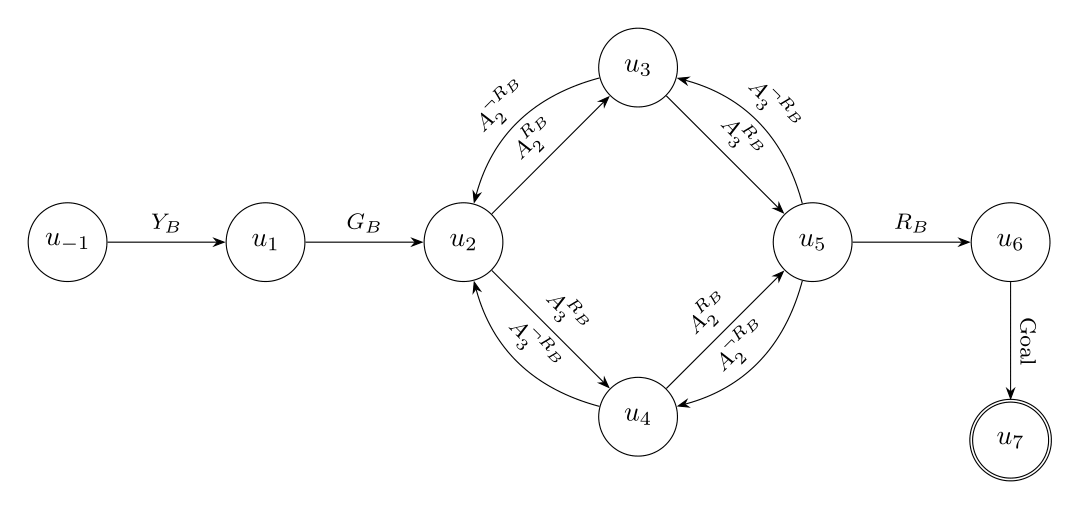}}  \\
        Overcooked & \makecell{\centering \includegraphics[width=0.5\textwidth]{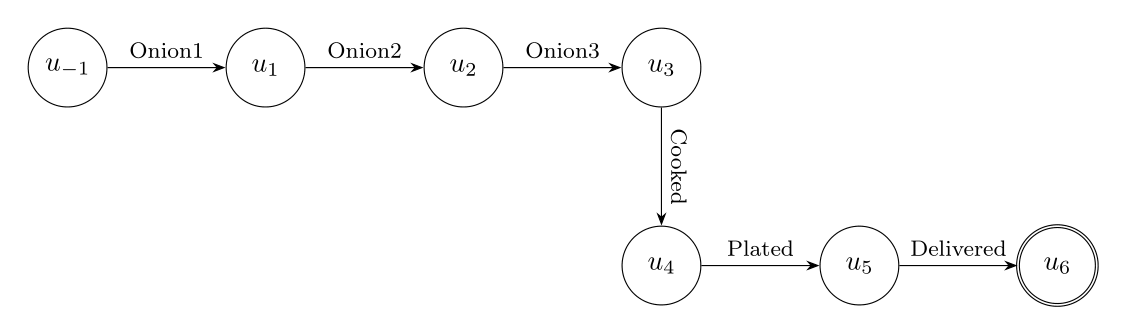}}  \\
        \hline
    \end{tabular}
    \caption{Reward machines for various environments.}
    \label{tab:reward_machines}
\end{table*}

\subsection{Hyperparameters and Environment Details}
\label{appendix:hyperparameters_env_details}
Table \ref{tab:hyperparameters} shows the PPO hyperparameters used for each training run in our experiments.




\paragraph{Cooperative Buttons} In the Cooperative Buttons task from~\cite{neary2020reward} used in our experiments, a goal region (denoted as `Goal') must be reached by any of the three agents. There are additionally three colored regions (red, yellow, and green), and three buttons of the same color present. In order for any agent to enter a colored region, the corresponding button must be pressed by at least one agent, otherwise, the attempt to enter that region results in a no-op. In the case of the red region, two agents must press the red button in order to enable an agent to traverse the red region. Walls (in black) separate Agent 1 from Agents 2 and 3. To reach the goal location, Agent 1 must first press the yellow button, which allows Agent 2 to traverse the yellow region to press the green button. Then, Agent 3 can traverse the green region, and both Agents 2 and 3 can press the red button, allowing Agent 1 to traverse the red region and reach the goal.

\paragraph{Four Buttons} In the Four-Buttons environment, a team of two agents must press the four buttons placed in the environment with an ordering constraint that the yellow button must be pressed prior to the red button.

\paragraph{Overcooked Environments} In the two Overcooked environments used in our experiments, the teams share the same task: three onions must be placed in the pot in order for the soup to start cooking. After a small number of timesteps have passed, the soup is considered cooked, and an agent must then pick up a plate and bring it to the pot in order to plate it. An agent then must take the plated soup and bring it to a delivery station. In the Asymmetric-Advantages environment, agents are in separate rooms, and both agents have shared access to a pot. Each agent has a distinct advantage (for example, the red agent can more easily place onions in the pot, whereas the blue agent can more quickly plate and deliver a cooked soup). In the Cramped-Corridor environment, only one agent may interact with the pot at a time, and the pot is placed in a small corridor. This can cause issues due to dependent agent dynamics: for example, if one agent finishes putting all of the onions in the pot, it must get out of the corridor either so the other agent can plate and deliver the soup.




\end{document}